\documentclass[]{spie}  

\usepackage{amsmath,amsfonts,amssymb}
\usepackage{graphicx}
\usepackage[colorlinks=true, allcolors=blue]{hyperref}

\title{SRAO: optical design and the dual-knife-edge WFS}

\author[a]{Carl Ziegler}
\author[a]{Nicholas M. Law}
\author[b]{Andrei Tokovinin}

\affil[a]{Department of Physics and Astronomy, University of North Carolina at Chapel Hill, Chapel Hill, NC 27599-3255, USA}
\affil[b]{Cerro Tololo Inter-American Observatory, Casilla 603, La Serena, Chile}

\authorinfo{Send correspondence to carlziegler@unc.edu}

\begin{document} 
\maketitle

\begin{abstract}
The Southern Robotic Adaptive Optics (SRAO) instrument will bring the proven high-efficiency capabilities of Robo-AO to the Southern-Hemisphere, providing the unique capability to image with high-angular-resolution thousands of targets per year across the entire sky. Deployed on the modern 4.1m SOAR telescope located on Cerro Tololo, the NGS AO system will use an innovative dual-knife-edge wavefront sensor, similar to a pyramid sensor, to enable guiding on targets down to V=16 with diffraction limited resolution in the NIR.  The dual-knife-edge wavefront sensor can be up to two orders of magnitude less costly than custom glass pyramids, with similar wavefront error sensitivity and minimal chromatic aberrations.  SRAO is capable of observing hundreds of targets a night through automation, allowing confirmation and characterization of the large number of exoplanets produced by current and future missions.
\end{abstract}

\keywords{natural guide star, adaptive optics, pyramid wavefront sensor, robotic telescope}

\section{INTRODUCTION}
\label{sec:intro}

The automation of adaptive optics observing, allowing unprecedented time-efficient observations, has been proven successful and worthwhile by the Robo-AO system \cite{riddle12, baranec13, baranec14}.  Expanding this capability to the larger SOAR telescope and providing access to the Southern-Hemisphere is the purview of the Southern Robotic Adaptive Optics (SRAO) instrument. In an era where exoplanet missions (e.g. \textit{Kepler}, TESS) find thousands of planetary candidate hosts stars, each of which requiring follow-up ground observations for confirmation and characterization, and ground-based wide-field surveys (e.g. Pan-STARRS, Evryscope, ZTF, LSST) produce multiple time-sensitive targets every night, the need for a robotic high-resolution instrument, capable of hundreds of observations a night, is evident.  Coupled with the already operational Northern Robo-AO system, and planned further Northern Hemisphere systems in Hawaii and elsewhere, all-sky robotic observations of up to 1000 targets a night will be possible.

Mounted on the SOAR telescope at the CTIO observatory, SRAO will take advantage of the excellent seeing and4m aperture size, along with a novel natural-guide-star photon-counting dual-knife-edge wavefront sensor (WFS) design, to enable guiding on targets down to V=15 with diffraction limited resolution (twice Hubble resolution).

In this paper, we discuss the technical design of SRAO, with an emphasis on the innovative dual knife-edge wavefront sensor.  In Section 2, we discuss the design system capabilities of SRAO.  In Section 3, the optical design and the robotic adaptive-optics observing software and image pipeline of the system is described.  In Section 4, we describe our novel design for wavefront sensing using knife-edge mirrors. We conclude in Section 6.

\section{SYSTEM CAPABILITIES}

SRAO will provide in the visible (650 nm) 0.03" FWHM imaging of sources V$<$10 with $\sim$20$\%$ Strehls over a 17" field-of-view.  In the infrared (J- and H-bands), SRAO will provide 0.08" FWHM on sources down to V=15, with 70$\%$ bright-target Strehls.  Expected typical overheads for the proven Robo-AO software are $\sim$1 minute with an LGS system.  With the simpler NGS system, we expect typical SRAO overheads to improve on Robo-AO, allowing observations of at least 10$\times$ more targets per hours than the similar MagAO system.

\begin{figure*}
\centering
\includegraphics[width=0.4\paperwidth]{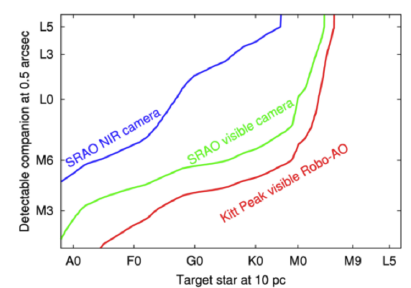}
\caption{Detectable companions around typical stars in SRAO multiplicity surveys.  The NIR camera is more effective at finding low-mass companions, and the visible camera will provide colors for mass estimates, improved angular resolution, and a passband that matches most large sky surveys.}
\label{fig:performance}
\end{figure*}

Using AO-assisted speckle-imaging, SRAO will achieve diffraction-limited visible-light performance on guide stars at least as faint as V=16, 1-2 magnitudes fainter than non-AO-assisted speckle imaging systems.  Compared to lucky imaging systems, SRAO will attain an increase in angular resolution of at least a factor of 2 and ten times more light-collection efficiency.  With an optional NIR camera upgrade path, detection of companions with 2-3$\times$ lower masses than other large-survey instruments is possible, including Robo-AO, as shown in Figure$~\ref{fig:performance}$.  A companion paper in these proceedings, Law et al. 2016, covers the science plans and capabilities of SRAO in more detail.  

\section{OPTICAL DESIGN}

\begin{figure*}
\centering
\includegraphics[width=0.7\paperwidth]{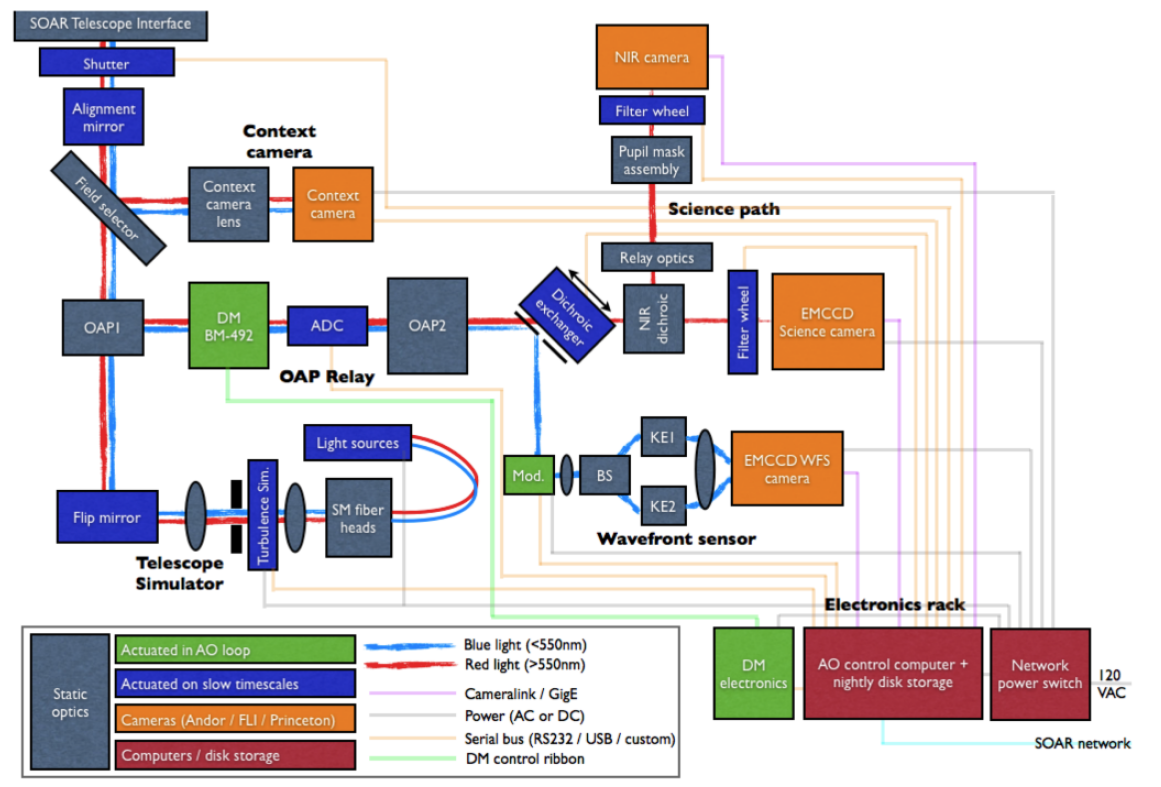}
\caption{Schematic of the major components of the SRAO system and their control paths.}
\label{fig:schematic}
\end{figure*}

The SOAR telescope has a 4.1m primary mirror.  SRAO will be mounted on the bent-Cassegrain port on the side of SOAR, taking in a beam with F/$\#$ of 16.63 and a plate scale of 3.025"/mm.  A target acquisition camera near the input will provide a larger seeing-limited field of view to enable automated target identification and alignment.  An OAP relay will provide magnification of 2.26$\times$, achieving Nyquist sampling for 30 mas visible-light diffraction-limited cores on detector.  An atmospheric dispersion corrector similar to Robo-AO's design is placed in the collimated beam after the DM and before the second OAP.  Visible light science images, with field-of-view approximately 17" square, are acquired by a photon-counting Andor iXon 888 EMCCD.  An optional upgrade path has NIR light sent by a dichroic into a re-imaging relay and to a Princeton Instruments 640LN NIR InGaAs-array camera.  The 640LN camera is liquid-nitrogen cooled, with significantly lower read noise and dark current (15e\textsuperscript{-} and $<$8e\textsuperscript{-}/pix/sec, respectively) than traditional off-the-shelf InGaAs cameras, allowing sky background limited observations in H-band.  Tip/tilt correction will be provided by SOAR's M3 rapid-actuation mirror.  The telescope simulator consists of a single-mode-fiber-fed collimated beam focused to the correct F$\#$ with rotating plastic disks to simulate turbulence.  Using a dichroic, part of the light will be sent from the science path to the WFS assembly.  The wavelengths extracted will be dynamically switched using an interchangeable dichroic assembly depending on the science goals.

\begin{figure*}
\centering
\includegraphics[width=0.5\paperwidth]{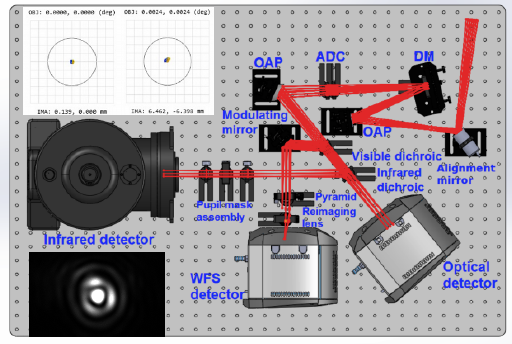}
\caption{SRAO optomechanical design showing major components and beam paths.  Light from the telescope enters at the top right.  In the top left corner inset: Zemax-simulated spot diagrams for on-axis images (left; Strehl ratio 0.99) and corner-of-field images(right; Strehl ratio 0.97).  The circle shows the diffraction-limited spot size for 656nm observations.  In the bottom left corner inset: PSF output of OAP relay from system testbed.}
\label{fig:design}
\end{figure*}

A schematic of the system is shown in Figure$~\ref{fig:schematic}$.  The mechanical design is shown in Figure$~\ref{fig:design}$.  The full SRAO Zemax model predicts, with perfectly-built and aligned optics and no atmosphere, center-of-field Strehl ratios at 656nm of 0.99, decreasing to 0.97 at the edge of the 17" field (Figure$~\ref{fig:design}$).

\subsection{Sofware Design}

The success of a robotic AO system is dependent on the reliability of its software design. SRAO will be built on the existing  Robo-AO software,\cite{riddle12} veteran of over 5 years of robotic AO development and observing. The Robo-AO control software autonomously operates Robo-AO's laser and safety systems, the adaptive optics control loop, the atmospheric dispersion corrector, and the science cameras. The system operates queue-scheduled, with autonomous optimal target selection and laser window avoidance.

The SRAO control software need only cover a smaller set of capabilities, as SRAO is initially planned to operate as a natural-guide-star system. The control software will be responsible for real-time wavefront reconstruction, DM control, tip/tilt removal, and queue based scheduling. Modifications of the Robo-AO code for use in SRAO include: 1) upgrades of the system performance for the 492-actuator system; 2) alteration of the system for natural guide star operation, a new reconstructor for the dual knife-edge WFS, interface with the Andor cameras and 640LN NIR camera, interface to SOAR TCS, and automatic acquisition of guide star with the context camera along with fast tip/tilt spiral slews coupled with fast frame rate science WFS EMCCDs. The Robo-AO reduction pipeline \cite{law14, ziegler15, ziegler16} will automatically calibrate and co-add the EMCCD visible-light camera data, and then perform automated PSF subtraction and companion detection.

\begin{figure*}
\centering
\includegraphics[width=0.7\paperwidth]{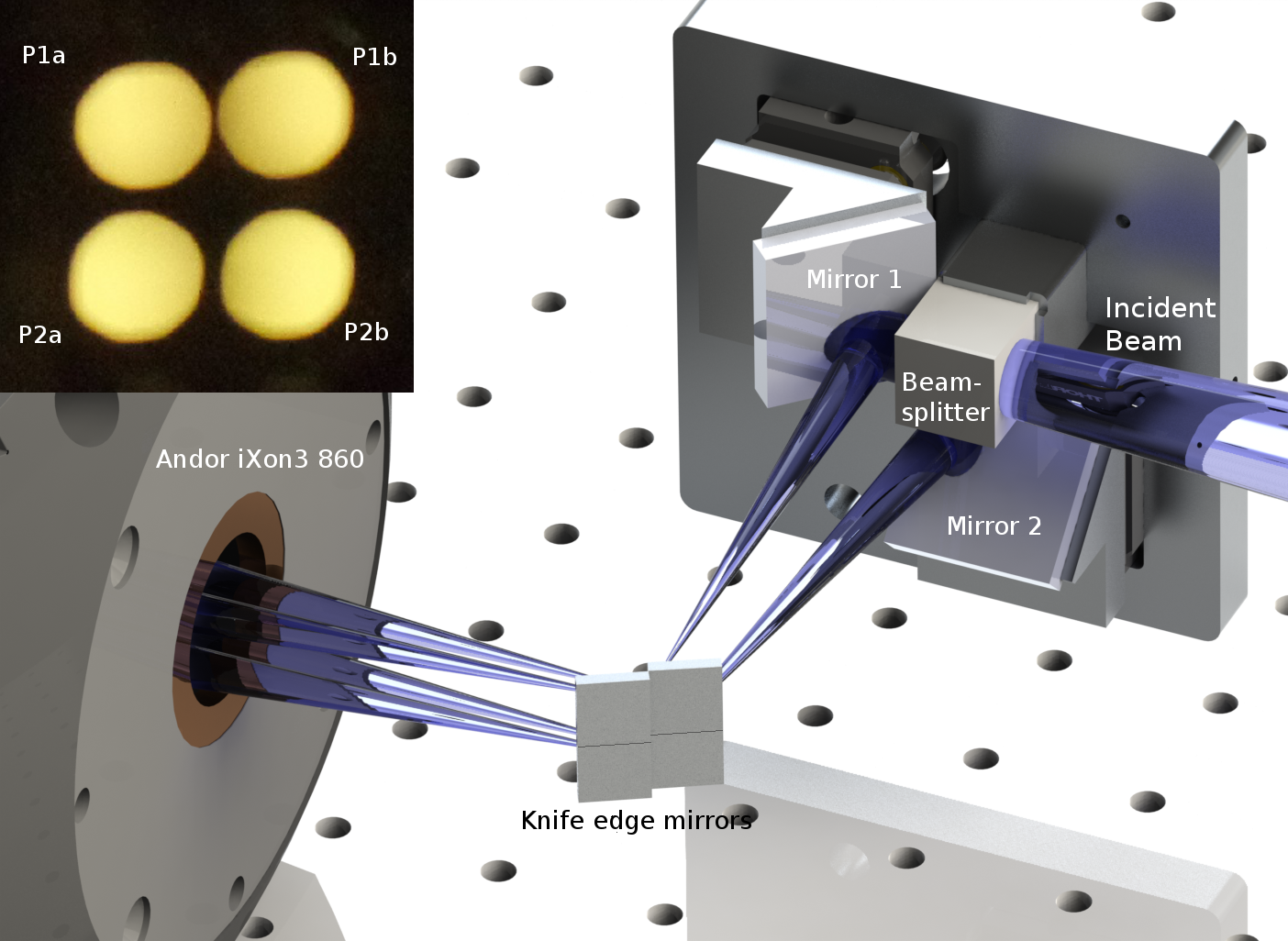}
\caption{Rendering of the dual knife-edge wavefront sensor.  In the upper left inset: four illuminated pupils output by the knife-edge wavefront sensor prototype with an input of diffuse light.}
\label{fig:wfs}
\end{figure*}

\section{DUAL KNIFE-EDGE WFS}

The SRAO WFS assembly is based on a pyramid-wavefront sensor (PWFS)\cite{riccardi98}, a system used on TNG, LBT, and Magellan, that has proven to effectively reach fainter guide-stars than Shack-Hartmann WFS systems\cite{chew06}.  The glass pyramid, placed at the focal plane of the beam, splits the light into four separate paths; a relay lens produces four images of the telescope pupil on the detector. Guiding on faint stars is then achieved by allowing dynamical rebinning of detector pixels to optimize the system for low-light levels, as well as taking advantage of AO image sharpening of the WFS images.  A single pyramid, however, suffers from severe chromatic aberrations, a problem that can be mitigated by employing a complex dual pyramid, as used for the LBT\cite{tozzi08}.  The expense of glass pyramids is a result of the precise requirements on their knife-edge vertices and base angles.

We have pioneered a new mostly-reflective system that removes the chromatic aberrations at greatly reduced cost and complexity.  The SRAO design uses crossed knife edges, with the light first divided by beamsplitter and then each image focused on a mirror with a knife-edge splitting the beam, forming two pupils for each slope direction.

Diffraction simulations performed using in-house custom IDL code demonstrate similar linearity range and response to tilts for a dual knife-edge sensor compared to a traditional PWFS, but with no cross-talk between the X and Y slope measurements and lower diffraction losses.  With the slopes sensed independently, as for a knife-edge sensor, the modulation required can be one-dimensional in each channel, thus reducing the cost and complexity of the required modulator.  For SRAO, modulation will be introduced before entering the WFS assembly by a PI S-316.10D tip/tilt steering mirror driven by a PI E-727.3SDA piezo-controller.

The design of the system is drawn in Figure$~\ref{fig:wfs}$, along with the illuminated pupil images from the WFS prototype.  The incoming light, redirected from the second OAP by the dichroic and increased to an F/100 beam with a large depth-of-focus using achromatic lenses, is split by a standard beamsplitter cube and directed to two plane mirrors resting on kinematic mounts which allow precise pointing of the beams.  Each beam is focused on separate sets of mirror pairs, made by diamond-cutting a mirror in the middle to get a sharp outer edge. In each set, one mirror is slightly behind the other, allowing the sharp edge of the leading mirror to divide the focal plane.  The edge of the back mirror is then in the shadow of the front, and thus only the edge quality of the leading mirror is important.  The mutual tilt of the two mirrors separates the reflected beams, and a pair of pupil images is formed with a re-imaging lens (not shown) onto the WFS CCD: pupil images P1a and P1b are formed from the upper knife-edge mirrors fed off of Mirror 1, and, likewise, P2a and P2b are formed off of the lower knife-edge mirrors fed off of Mirror 2.  With the 90$^{\circ}$ twist between the beams introduced by the beamsplitter and Mirrors 1 and 2, the parallel knife-edge mirrors are able to sense both the X and Y tilts of the incoming wavefront.  The difference in intensity between P1a and P1b serves to sense the X-tilt, and the pair P2a and P2b to sense the Y-tilt.  This also allows a one-dimensional modulator to be used, reducing both the cost and complexity of the system.

\begin{figure*}
\centering
\includegraphics[width=0.7\paperwidth]{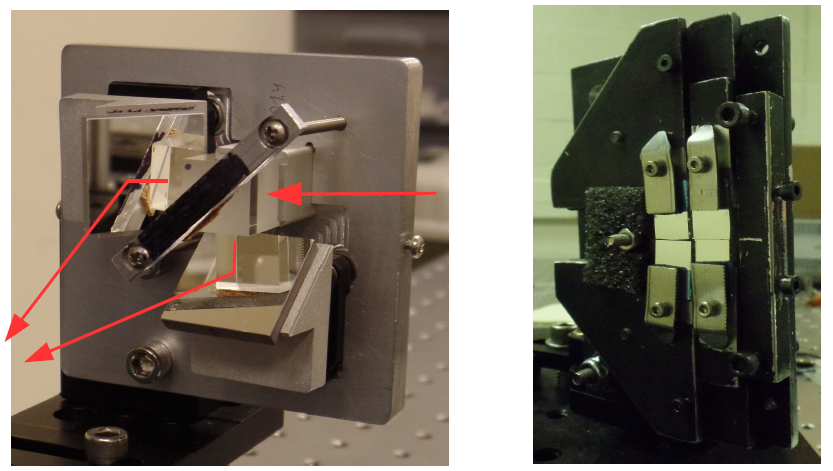}
\caption{The prototype dual knife-edge wavefront sensor: the beamsplitter module (left), and the knife-edge mirror module (right).}
\label{fig:prototype}
\end{figure*}

\subsection{Wavefront Sensor Prototype}

The prototype of the wavefront sensor assembly is shown in Figure$~\ref{fig:prototype}$.  On the left, the incoming light (red arrow from right of image) passes through the beamsplitter cube, and both result rays are reflected toward the knife-edge module (right).  The two beams cross, such that the lower beam is focused on the upper mirror pair, and the upper beam on the lower mirror pair.  The mirrors in the beamsplitter module, 1-inch square and silver-coated, are mounted on kinematic mirror mounts by means of the 45-degree machined bases for tip-tilt adjustment, allowing fine placement of the resulting beams on the knife-edge of the mirrors.  The knife-edge module also allows tip-tilt adjustment of each pair individually as well as global tilt adjustment of the pairs with respect to each other.  The resulting pupils can then be precisely placed on the detector in each quadrant in a square arrangement, shown in Figure$~\ref{fig:wfs}$.  A lab testebed performance characterization of the WFS prototype is ongoing.

\section{CONCLUSION}

SRAO will provide automated, moderate-order NGS AO to the 4.1m SOAR telescope at CTIO.  Along with the Northern Hemisphere Robo-AO and planned Hawaii and elsewhere systems, all-sky high-efficiency AO observations of up to 1000 targets a night will be possible, allowing large, previously infeasible surveys to be performed.  With an innovative dual knife-edge WFS, similar in concept to a pyramid WFS but with reduced chromatic aberrations, SRAO can reach the diffraction limit on brighter targets.  SRAO is currently in testbed phase, with plans to deploy as a visitor instrument on SOAR in 2017.

\bibliography{main.bib} 
\bibliographystyle{spiebib} 

\end{document}